National Scientific Facilities and Their Science Impact on Non-Biomedical Research

A.L. Kinney, NASA/GSFC, 8800 Greenbelt Road, Greenbelt, MD 20771

Abstract

H-index, proposed by J.E. Hirsch (1) is a good indicator of the impact of a scientist's research and has the advantage of being objective.   When evaluating departments, institutions or labs, the importance of h-index can be further enhanced when properly calibrated for size of the group.  Particularly acute is the issue of federally funded facilities whose number of actively publishing scientists frequently dwarfs that of academic departments.   Recently Molinari & Molinari (2) developed a methodology that shows the h-index has a universal growth rate for large numbers of papers, allowing for meaningful comparisons between institutions.

An additional challenge when comparing large institutions is that fields have distinct internal cultures, with different typical rates of publication and citation; biology is more highly cited than physics, which is more highly cited than engineering.   For this reason, this study has focused on the physical sciences, engineering, and technology, and has excluded bio-medical research.   Comparisons between individual disciplines are reported here to provide contextual framework.  Generally, it was found that the universal growth rate of Molinari & Molinari holds well across all the categories considered, testifying to the robustness of both their growth law and our results.

The overall goal here is to set the highest standard of comparison for federal investment in science; comparisons are made with the nations preeminent private and public institutions.  We find that many among the national facilities compare favorably in research impact with the nations leading universities.

Introduction

The "h-index", pioneered by Hirsch (1), has rapidly become a widely used marker for evaluating impact of scientific research.  The h-index of an individual scientist is defined as the number of his/her publications cited more than h times in scientific literature.  Similarly, h-index can be generalized to groups of scientists, departments, and large institutions.  Recently, Molinari and Molinari (2, M&M) have observed that, when evaluating sets of publications that are greater than several hundred, h-index versus the size of the set of publications (N) is characterized by an approximately universal growth rate, referred to by M&M as the master curve.    The underlying reason for this is a topic for another paper, and may have to do with the speed of the diffusion of knowledge and an intrinsically non-linear relation between number of publications and h-index.  Regardless, the observational fact that there is such a universal growth rate allows h-index to be decomposed into the product of an impact index and a factor depending on the size of the set, which in turn allows for a meaningful comparison between institutions of widely varying sizes.  The growth rate is given by $N^{**}0.4$, where N is the total number of papers, so that the impact index defined by M&M is given by h(m) = h-index / N**0.4.

Impact Index as a function of Scientific Discipline

To demonstrate this universal growth rate, and to better understand the differences between the scientific disciplines, the total number of papers and corresponding h-indices have been assembled here using Web of Science (3) according to the disciplines which are encompassed by this study. Both small fields like astronomy are included as are large fields, like physics, mathematics, and chemistry.

Any search involving Web of Science has intrinsic limitations due to the nature of their search engine. This particular search was done based on Departments in the United States. There are some ambiguities in such a data search. For example, some Universities have physics and astronomy within one department. The astronomy-related publications from such a physics department would not be included in a search for publications from Departments of Astronomy. While the higher citation rate of astronomy would increase the numbers for such a physics department, there are many fewer astronomers than physicists, so in the end, there is little effect on the final numbers. Given the universality of the growth rate, the astronomy publications not included because of their affiliation with a physics department do not have much influence on the numbers for the field of astronomy, making them somewhat lower, but not changing their impact when normalized by size.

Once the publications are gathered in ISI, the set is scrutinized for bio-medical and other unrelated publications. Fields related to health science, biology, biophysics, medicine, the human body, diseases, social science, agriculture, and other unrelated fields were removed. The disciplines that are represented in this study were left in the sample. Note that the topic of the specific department dominates the total number of papers so that leaving the other physical sciences and engineering/technology disciplines in the set of publications has a minor effect on the overall numbers. For the mathematical discipline, the search found 22,261 publications under the subject category "mathematics", 11,950 under applied mathematics, an additional 3,066 under statistics and physics-related mathematics while the first field that was not directly mathematical was multidisciplinary physics with 1,275 publications. See the Appendix for further details of the methodology.

One major way in which this study differs from that of M&M is that publications were "innocent until proven guilty" --- that is, they were not removed unless they were specifically identified as irrelevant. This method results in a larger range of science topics being included, such as geosciences, oceanography, earth science, computer science, and especially engineering and technology development so as to better make the comparison with the federal science centers and laboratories of NASA, DOE, and NSF, where research and development in these areas is a significant component of the overall output. The number of publications included in this study is thus approximately three times the number included in the study of M&M.

Once the set is scrutinized for bio-medical publications and other unrelated topics, the growth rate is calculated following the methodology of M&M by first producing the h-index for the publications in the year 1980, then for the years 1980 plus 1982, and so on until the h-index for the accumulated publications from 1980 to 1998 are calculated. Publications after 1998 are not included because those publications have not had the time to be fully cited. The lack of full citation is manifested by the time dependence of the impact indices, and can be seen in Figure 4 of M&M.

When the h-index and total numbers of papers are plotted on a log-log plot, this universal growth rate is manifest, following a curve of slope approximately 0.4, as seen in Figure 1. The scientific fields included in this study tend to fall under two groupings; astronomy, physics, and chemistry have high characteristic impact indices (5.14, 4.01, and 3.52, respectively), while engineering, mechanical engineering, and mathematics have smaller characteristic impact indices (2.79, 2.60, and 2.33 respectively, see also Table 1). In Figure 1, two template lines with slope of 0.4 are drawn encompassing the master curves of these six disciplines, which will also be placed on subsequent figures as a guide. These master curves represent a template for academic impact in the United States.

Table 1: Impact Index according to Scientific Disciplines

| Discipline | N | h-index | h(m) |
|---|---|---|---|
| Astronomy | 7459 | 182 | 5.14 |
| Physics | 80081 | 367 | 4.01 |
| Chemistry | 144183 | 408 | 3.52 |
| Engineering | 3706 | 74 | 2.79 |
| Mechanical Engineering | 16088 | 125 | 2.60 |
| Mathematics | 40253 | 162 | 2.33 |

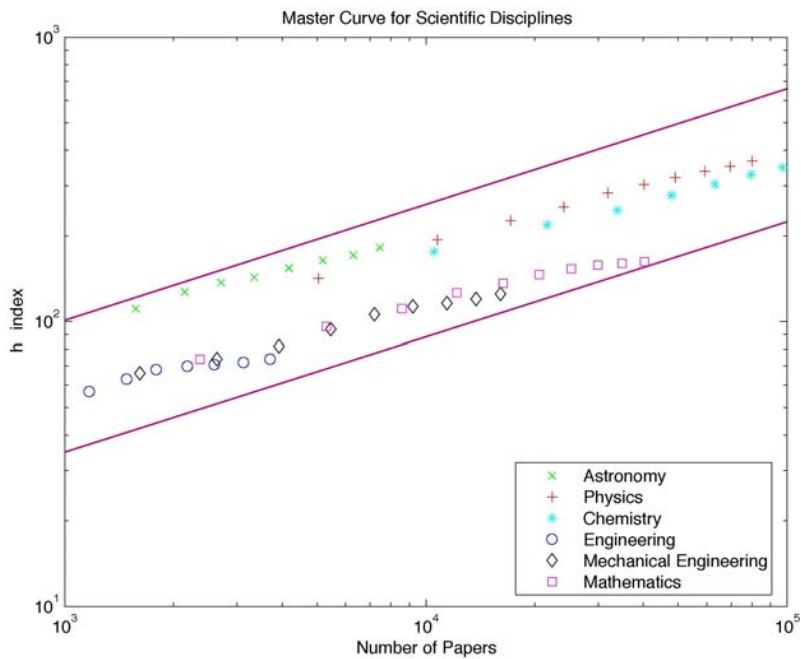

Figure 1: Master Curve for Science Disciplines. The h-index is calculated for non-biomedical publications for the year 1980, 1980 plus 1982, incrementing the years included by even years through 1990. Overlaid are lines with slope 0.4. While the universal law with exponent 0.4 works well for physics, chemistry, and astronomy, it works less well for engineering, and mathematics, where a somewhat lower exponent of the order of 0.35 might be more appropriate.

Top Ranked American Academic Institutions

To further establish context for an evaluation of federally funded science centers, the impact indices for the top ranked American academic institutions are presented here. These present a different challenge; the publications of many of the top ranked institutions are dominated by bio-medical research, which dwarf the other scientific disciplines in the number of published papers, journals, annual reports, and refereed proceedings. Here we define the "panning for gold" approach; Publications in approximately 520 highly utilized bio-medical, medical and health related journals were systematically excluded in an "advanced search" in ISI Web of Science (3). The same list of journals was used in every search. The resulting set of papers was then weeded by hand for subject categories involving bio-medical, medical, health, and other unrelated topics. The master curve was then composed of the data for 1980, then 1980 plus 1982, adding one even year of publications until all the even years from 1980 to 1998 were included. All of the details of the search including journal titles excluded, example subject categories excluded by hand, and syntax used are given in the Supporting Information. This approach was to "pan for gold" in that the many undesired bio-medical publications were removed from the few desired science and technology publications so as to maintain a broad spectrum of disciplines in the set.

This method was used for both the top ranked American Institutions and also for the Big Ten universities so as to get a broad characterization of preeminent public and private US academic impact indices.   The master curves for these institutions are shown in Figures 2 and 3, with the template discipline master curves included for reference.   Table 2 shows the impact indices for some of the top US Universities. Compared to M&M's list of a similar nature,  Harvard and Princeton are still in the top third, Stanford, Chicago, and Columbia are still in the middle, and Duke and UC Berkeley are in the bottom third. JHU has gone from eighth to second, CALTECH from second to sixth, and MIT from forth to ninth.

Clearly evaluations of impact are dependent on the disciplines included in the study.  This study covers a broader range of sciences and technologies, with three times as many publications included as the study of M&M, with; both of these factors no doubt influence the final numbers.

Table 2: Impact Index of Top Ranked U.S. Universities

| Top Ranked Universities | N | h-index | h(m) |
| --- | --- | --- | --- |
| Harvard University | 11165 | 256 | 6.15 |
| Johns Hopkins University | 5959 | 167 | 5.16 |
| Princeton University | 9084 | 197 | 5.14 |
| Columbia University | 7028 | 174 | 5.03 |
| University of Chicago | 6354 | 167 | 5.03 |
| CALTECH | 13381 | 217 | 4.85 |
| Stanford University | 13215 | 213 | 4.79 |
| Duke University | 3724 | 123 | 4.74 |
| MIT | 19542 | 241 | 4.63 |
| UC Berkeley | 19963 | 220 | 4.19 |

Figure 2 shows the master curve for top ranked universities, including template lines carried over from Figure 1.  The h-index is calculated for non-biomedical publications for the year 1980, 1980 plus 1982, incrementing the years included by even years through 1990.  See Supporting Information for detailed methodology.

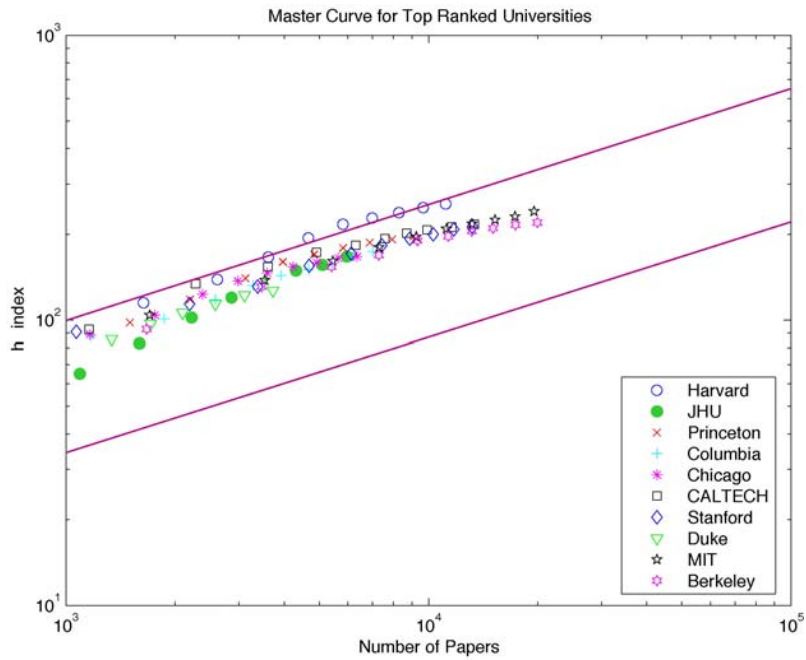

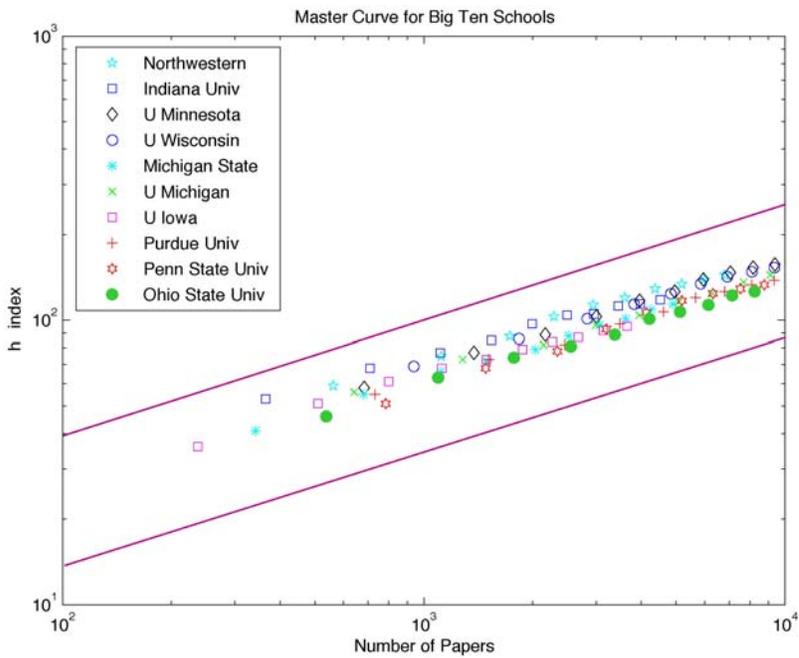

Figure 3 shows the master curves for the big ten Universities. The most remarkable aspect of the impact index for the Big Ten universities is the narrow dispersion of the curves and the impact indices, shown in Table 3.

Table 3: Impact Index for Big Ten Universities

| Big Ten Universities | N | h-index | h(m) |
|---|---|---|---|
| Northwestern Univ | 6801 | 144 | 4.22 |
| Indiana University, | 4518 | 118 | 4.07 |
| Bloomington | | | |
| Univ of Minnesota, St Paul, Minneapolis, & Twin Cities | 9370 | 158 | 4.05 |
| Univ of Wisconsin, Madison | 10647 | 156 | 3.81 |
| Michigan State | 4894 | 114 | 3.81 |
| Univ of Michigan, Ann Arbor | 10657 | 150 | 3.67 |
| Univ of Iowa, Iowa City | 3604 | 95 | 3.59 |
| Purdue Univ | 10582 | 141 | 3.46 |
| Penn State Univ, University Park | 10170 | 138 | 3.44 |
| Ohio State Univ, Columbus | 8230 | 126 | 3.42 |

NASA Science Centers

The data for the NASA science centers Goddard Space Flight Center, Ames Research Center, Marshall Space Flight Center, the Federally Funded Research and Development Center, Jet Propulsion Laboratory, and Langley Research Center are presented. All of these NASA centers have a great enough output of publications so as to fall onto the universal growth curve of M&M.

The different NASA centers have traditionally focused on different areas of interest. Of greatest impact to an evaluation of their science impact, however, is the percentage of their publications that concern scientific topics as opposed to the percentage that concern technology and engineering, especially taking into account the gap between the impact indices of astronomy, physics and chemistry versus the citation rates of engineering as reported above. Using the Web of Knowledge (3), the publications from these centers have been put into two categories according to their subject category; the first category is broadly termed "science" and includes astronomy, meteorology, geosciences, physics, planetary science, earth science, oceanography, and chemistry. The second category is broadly termed "engineering", and includes topics related to the research and development of new technologies, engineering, remote sensing, optics, computer science, telecommunications, robotics, and applied sciences. Table 1 lists the NASA Centers and the percentage of their publications according to whether they are science or engineering according to this definition, along with their impact indices.

Table 4: Impact Index for NASA Centers

| NASA Facilities | % Science | % Engineering | N | h-index | h(m)= h/N**0.4 |
|---|---|---|---|---|---|
| Goddard Space Flight | 78% | 22% | 6300 | 144 | 4.35 |

| Center | | | | | |
|---|---|---|---|---|---|
| Ames Research Center | 70% | 30% | 2803 | 123 | 5.14 |
| Marshall Space Flight Center | 64% | 36% | 1320 | 70 | 3.95 |
| Jet Propulsion Lab | 55% | 45% | 5183 | 128 | 4.18 |
| Langley Research Center | 28% | 72% | 2948 | 86 | 3.52 |

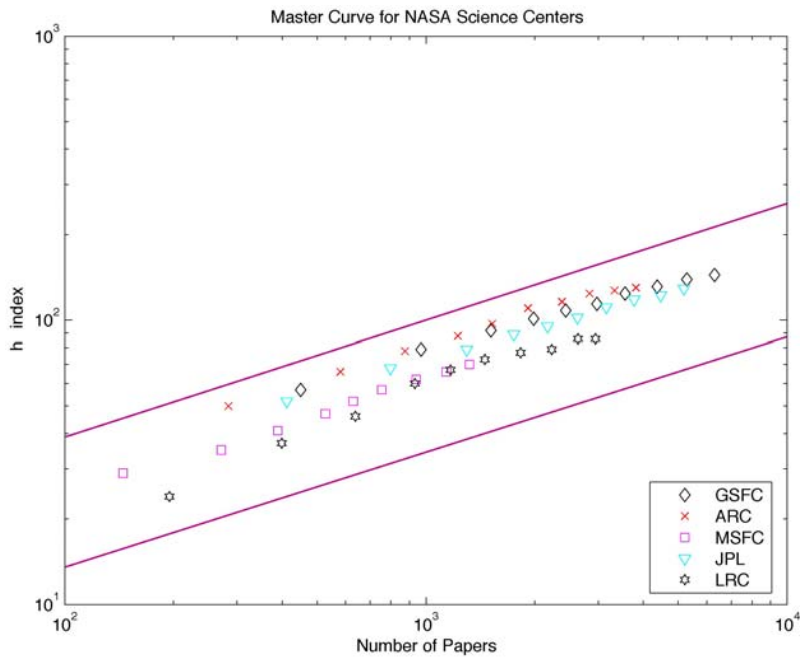

Figure 4, above, shows the master curves for these five centers together with the template master curves from in Figure 1. The impact index for the NASA centers approximately reflects the percentage of science publications versus engineering publications, and falls within the template master curves for the six science disciplines.

NASA Centers compare very favorably with the Big Ten universities despite the substantial engineering/technology component of the NASA centers.

Department of Energy

The master curves for the DOE science centers Stanford Linear Accelerator Laboratory, Fermi National Accelerator Laboratory, Brookhaven National Laboratory, Lawrence Livermore National Laboratory, Lawrence Berkeley National Laboratory, Argonne National Laboratory, Los Alamos National Laboratory, and Oak Ridge National Laboratory, are shown in Figure 5 together with CERN as a comparison and with the template lines from Figure 1. The data fall into two categories, with SLAC, Fermi Lab, Brookhaven, LLNL and LBNL in the higher range, and Argonne, Los Alamos, and Oak Ridge in the lower range. Naively, since these data come from a highly international community, where papers have authorship that number in the tens if not hundreds of authors from many institutions, the citation rates would be expected to be more tightly clustered than they are.

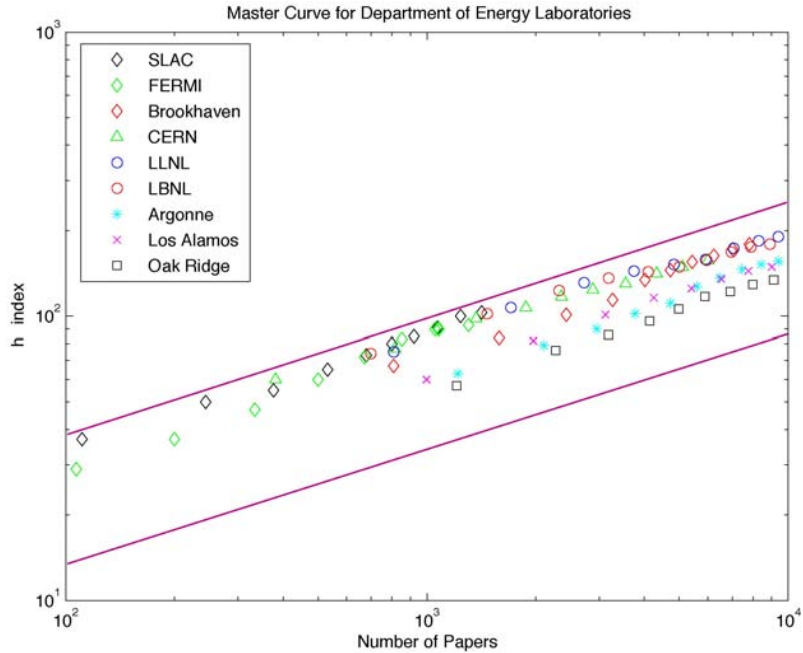

Table 5: Impact Index for DOE National Labs + CERN

| DOE Laboratories | N | h-index | h(m) |
|---|---|---|---|
| SLAC | 1418 | 103 | 5.65 |
| Fermi Nat Acc Lab | 1304 | 93 | 5.28 |
| Brookhaven NL | 7809 | 179 | 4.96 |
| CERN | 5999 | 157 | 4.84 |
| LLNL | 10605 | 196 | 4.81 |
| LBNL | 8900 | 179 | 4.71 |
| Argonne NL | 9413 | 156 | 4.01 |
| Los Alamos NL | 11776 | 163 | 3.83 |
| Oak Ridge NL | 10266 | 138 | 3.43 |

National Science Foundation and Astronomical Observatories

Most of the facilities and centers supported by the National Science Foundation in physics, mathematics, engineering, mechanical engineering and geosciences cannot be analyzed by this method because of their distributed nature. NSF facilities are often distributed amongst a consortium of members, as opposed to being long term facilities which are identified in the address of a publication. Since there is no way to identify the publications with the particular center, publication and citation statistics cannot be collected. Astronomy facilities are the exception to this rule, with National Radio

Astronomy Observatory, National Optical Astronomical Observatory, National Solar Observatory, all having fixed addresses (sometimes in more than one location) and having been scientifically active for long enough to evaluate. The Gemini Observatory is an additional NSF astronomical facility with an identifiable address for publication searches, but it is too new to lie on the uniform growth curve of the master curve. There are also two physics centers supported by NSF which can be evaluated here, both with a contiguous publication presence with high enough numbers to make a valid assessment of impact index; Kavli Institute of Theoretical Physics at UC Santa Barbara, and National High Magnetic Field Laboratory. Although both institutes represent a different discipline than astronomy, they are NSF funded institutes with comparable lifetime of operations as the other institutes evaluated here and so are included in the discussion. KITP is also included in the master curve, while National High Magnetic Field Laboratory does not have a long enough time line to analyze in that way.

The impact index was calculated differently for these cases than for the other institutes included in this study so as to be able to make a better comparison between operating years since some institutions are newer and were not publishing in the early 1980's. The 9 years from 1990 to 1998 were included in the calculation of h-index. For three of the institutions, ITP, NOAO, and NRAO, there were enough publications over a long enough time scale to compose a master curve in the same manner as for all other institutions presented here, i.e. utilizing data from even years between 1980 and 1998. For STScI, there are no publications for 1980, when it was begun, and for NOAO, there is no data for 1980 and 1982. The master curve is shown in Figure 6. The NASA funded center, Space Telescope Science Institute is included given its closely related science. Also included for comparison is the impact index for the Astronomy Department, UC Berkeley.

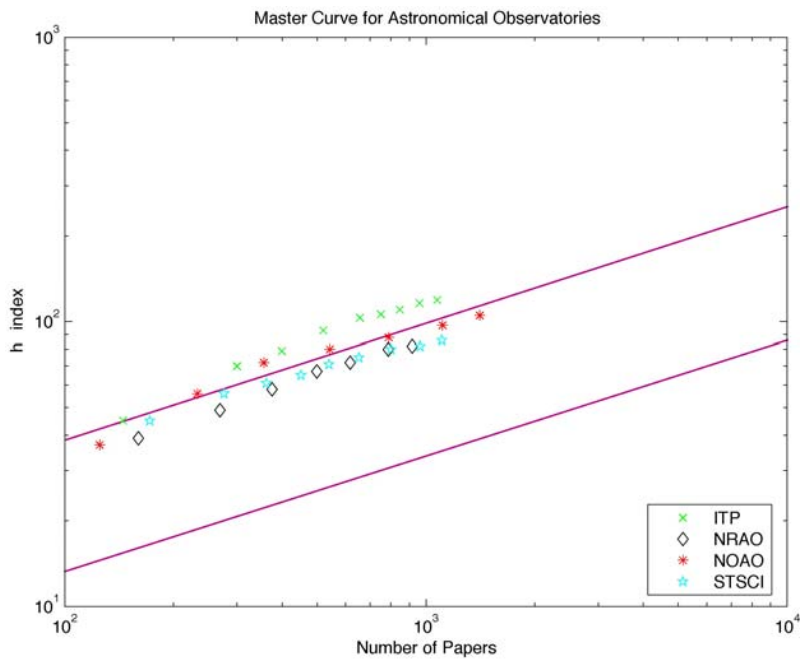

Based on the master curves shown in Figure 6, above, the NSF observatories, ITP and STScI all have high impact indices. Since these institutes represent single disciplines, it is not appropriate to compare them with the institutions which encompass much broader scientific disciplines.

Table 6: Impact Index for NSF Facilities plus STScI

| Astronomical Institutions | N | h-index | h(m) |
| --- | --- | --- | --- |
| KITP, UCSB* | 972 | 103 | 6.56 |
| UC Berkeley Astronomy | 1241 | 109 | 6.30 |
| NOAO | 1133 | 90 | 5.40 |
| STScI | 2161 | 116 | 5.38 |
| National Solar Obs. | 333 | 43 | 4.21 |
| Nat High Mag Field Lab | 940 | 62 | 4.01 |
| NRAO | 2122 | 80 | 3.74 |

*Note that KITP is a theoretical physics institute which regularly organizes conferences and long term workshops in physical sciences. Many of the authors of papers with KITP affiliation are visitors, with other home institutions.

Comparisons and Conclusions

Here, a number of federally funded science centers and laboratories have been compared with the highest ranking public and private academic institutions. An overall comparison is shown in Figure 7, with data from each type of institution displayed with a different icon. The top ranked academic institutions in the United States have the highest impact index among all the institutions evaluated here, followed by leading DOE Laboratories and NASA Centers that generally rank higher than the Big Ten universities.

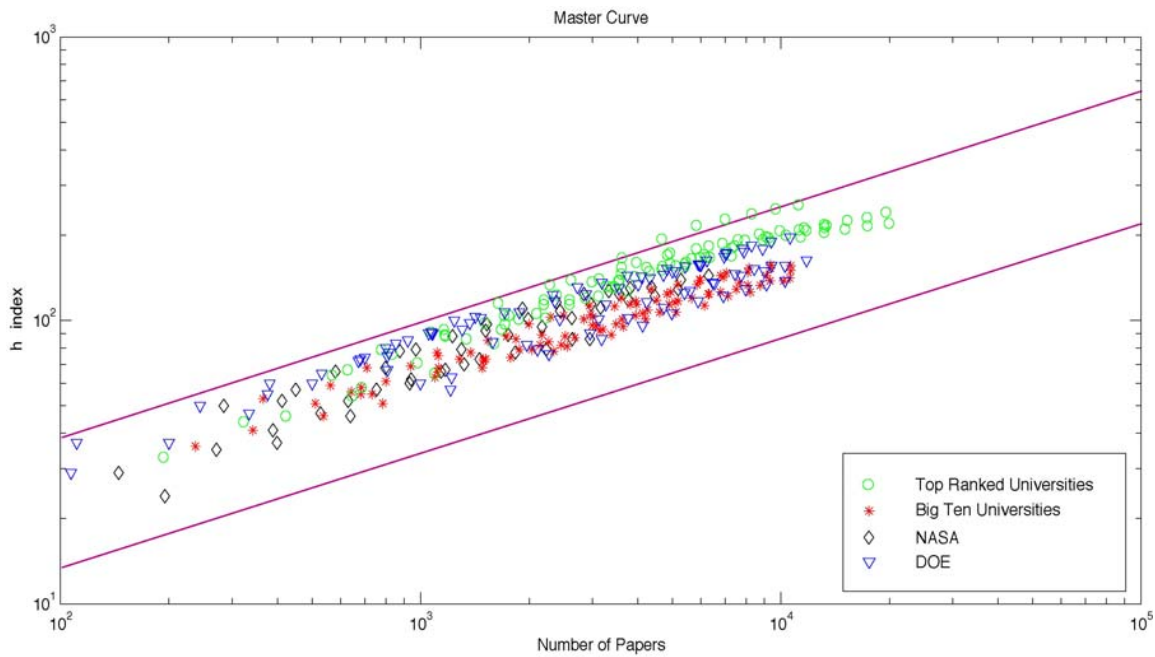

References

1. Hirsch, J.E. (2005) Proc. Nat. Acad. Sci., **102**, 16569-16573.
2. Molinari J.F. & Molinari, A. (2007) submitted to Scientometrics.
3. ISI Web of Knowledge, http://portal.isiknowledge.com

SUPPORTING INFORMATION

Methodology

For the compilation of the set of publications according to science discipline, the "advanced search" option in ISI Web of Science (3) was used to select by discipline and by year of publication.    For the creation of the science discipline master curves for Departments of Mathematics, for example, the advanced search command used was:

AD=dept math SAME AD=usa and (PY=1980 or PY=1982 or PY=1984 or PY=1986 or PY=1988 or PY=1990 or PY=1992 or PY=1994 or PY=1996 or PY=1998).

The set of publications was then weeded according to "subject category". When collating the publications for "Dept Math", the following specific subject categories were excluded; agriculture – diary & animal science, agriculture – multidisciplinary, agriculture – soil science, agronomy, anatomy & morphology,  anesthesiology, applied linguistics, behavioral science, biochemical research methods, biochemistry & molecular biology, biodiversity – conservation, biology, biophysics, biotechnology & applied microbiology, business, business – finance, cardiac & cardiovascular systems, cell biology, chemistry – medicinal, clinical neurology, dentistry – oral surgery & medicine, dermatology, developmental biology, endocrinology & metabolism, engineering - biomedical, entomology, ergonomics, ethics, evolutionary biology, forestry, food science

& technology, gastroenterology & hepatology, genetics & heredity, geriatrics & gerontology, health care sciences & services, health policy & services, hematology, immunology, infectious diseases, medical laboratory technology, medicine – general & internal, medicine – informatics, medicine – legal, medicine – research and experimental, microbiology, neurosciences, nuclear medicine & medical imaging, obstetrics & gynecology, ornithology, nutrition & dietetics, oncology, ophthalmology, orthodepdics, otorhinolaryngology, pathology, pediatrics, peripheral vascular disease, pharmacology & pharmacy, physiology, plant sciences, psychiatry, psychology – biological, psychology – multidisciplinary, public, environmental & occupational health, radiology, respiratory systems, rheumatology, social issues, social science – mathematical methods, sports sciences, substance abuse, surgery, toxicology, urology & mephrology, veterinary sciences, virology, and zoology.

The master curve was then created using first the data from 1980, then 1980 plus 1982, and so on until the even years from 1980 to 1998 were included.

For the compilation of the set of publications for the top ranked and Big Ten universities, the advanced search excluded bio-medical and health related journals and undesired topics. The advanced search option is limited to 50 boolean operators, and so the approximately 520 undesirable titles and 80 undesirable topics were broken up into 16 individual searches. The searches included only the 10 years used to produce the growth curve in order to remain under the 100,000 limit of Web of Science searches. The 16 resulting lists of exclusions were then combined with the Boolean operator "and" so as to find the publications common to all searches. This list was then searched by hand according to subject category.

A preferable method would have been to include subject category in the advanced search so as to have a uniform search not subject to the human error of an individual search. Unfortunately, subject category cannot be included as a "field tag", i.e. a search criteria, in an advanced search, and so this part of the exclusion of bio-medical and other non-related publications was done by hand. Instead, the field tag for topic (TS) was used in the advanced search to exclude bio-medical topics. The field tag "topic" or "TS" searches the abstract, and excludes publications which refer to the identified topic in the abstract. For this reason topics like "history", which can have a general usage in an abstract, i.e. "the history of the universe", must not be used or they will result in the exclusion of relevant publications.

A typical list of excluded subject categories is given at the bottom, and includes bio-medical subjects as well as such diverse topics as agriculture, anthropology, horticulture, genetics & heredity, and paleontology   The weeding of these subject categories was done until single article categories were being removed.

#1

AD=INDIANA UNIV and (PY=1980 or PY=1982 or PY=1984 or PY=1986 or PY=1988 or PY=1990 or PY=1992 or PY=1994 or PY=1996 or PY=1998) not (SO=JOURNAL OF BIOLOGICAL CHEMISTRY or SO=FASEB JOURNAL or SO=CIRCULATION or SO=

INVESTIGATIVE OPHTHALMOLOGY & VISUAL SCIENCE or SO=BIOPHYSICAL JOURNAL or SO=CLINICAL RESEARCH or SO=FEDERATION PROCEEDINGS or SO=JOURNAL OF CELL BIOLOGY or SO=JAMA-JOURNAL OF THE AMERICAN MEDICAL ASSOCIATION or SO=NEW ENGLAND JOURNAL OF MEDICINE or SO=GASTROENTEROLOGY or SO=CANCER RESEARCH or SO= BLOOD or SO=AMERICAN JOURNAL OF EPIDEMIOLOGY or SO=BIOCHEMISTRY or SO=JOURNAL OF UROLOGY or SO=ARCHIVES OF OPHTHALMOLOGY or SO= JOURNAL OF IMMUNOLOGY or SO= PEDIATRIC RESEARCH or SO=(JOURNAL OF ALLERGY AND CLINICAL IMMUNOLOGY) or SO= PEDIATRICS or SO=MOLECULAR BIOLOGY OF THE CELL or SO=(JOURNAL OF ALLERGY AND CLINICAL IMMUNOLOGY) or SO= JOURNAL OF INFECTIOUS DISEASES or SO= LABORATORY INVESTIGATION or SO= JOURNAL OF VIROLOGY or SO= LANCET or SO= OPHTHALMOLOGY or SO= ANNALS OF NEUROLOGY or SO= BRAIN RESEARCH or SO= AMERICAN JOURNAL OF PHYSIOLOGY or SO=AMERICAN JOURNAL OF PUBLIC HEALTH or SO= JOURNAL OF PEDIATRICS or SO= (MOLECULAR AND CELLULAR BIOLOGY) or SO= (ARTHRITIS AND RHEUMATISM) or SO= AMERICAN JOURNAL OF MEDICAL GENETICS or SO= AMERICAN JOURNAL OF OPHTHALMOLOGY or SO= JOURNAL OF NEUROSCIENCE or SO= AMERICAN JOURNAL OF HUMAN GENETICS)

#2

AD=INDIANA UNIV and (PY=1980 or PY=1982 or PY=1984 or PY=1986 or PY=1988 or PY=1990 or PY=1992 or PY=1994 or PY=1996 or PY=1998) not (SO= AMERICAN REVIEW OF RESPIRATORY DISEASE or SO= NEUROLOGY or SO= ANNALS OF INTERNAL MEDICINE or SO= EXPERIMENTAL HEMATOLOGY or SO=AMERICAN JOURNAL OF MEDICINE or SO=ANESTHESIOLOGY or SO=HEPATOLOGY or SO= JOURNAL OF CELLULAR BIOCHEMISTRY or SO=JOURNAL OF NEUROCHEMISTRY or SO= CANCER or SO= JOURNAL OF INVESTIGATIVE DERMATOLOGY or SO= (BIOCHEMICAL AND BIOPHYSICAL RESEARCH COMMUNICATIONS) or SO= CELL or SO= AMERICAN JOURNAL OF CLINICAL NUTRITION or SO= JOURNAL OF CLINICAL INVESTIGATION or SO= JOURNAL OF THE AMERICAN GERIATRICS SOCIETY or SO= JOHNS HOPKINS MEDICAL JOURNAL or SO=JOURNAL OF APPLIED PHYSIOLOGY or SO= NUCLEIC ACIDS RESEARCH or SO= PROCEEDINGS OF THE AMERICAN ASSOCIATION FOR CANCER RESEARCH or SO= (JOURNAL OF NEUROPATHOLOGY AND EXPERIMENTAL NEUROLOGY) or SO= (AMERICAN JOURNAL OF TROPICAL MEDICINE AND HYGIENE) or SO= JOURNAL OF CLINICAL MICROBIOLOGY or SO= GENOMICS or SO= (JOURNAL OF PHARMACOLOGY AND EXPERIMENTAL THERAPEUTICS) or SO= ENDOCRINOLOGY or SO= AMERICAN JOURNAL OF CARDIOLOGY or SO=JOURNAL OF THE AMERICAN SOCIETY OF NEPHROLOGY or SO= (OBSTETRICS AND GYNECOLOGY) or SO= STROKE or SO= KIDNEY INTERNATIONAL or SO= RADIOLOGY or SO=(JOURNAL OF CLINICAL ENDOCRINOLOGY AND METABOLISM) or SO= JOURNAL OF THE NATIONAL CANCER INSTITUTE or SO= (JOURNAL OF NERVOUS AND MENTAL DISEASE) or SO= TRANSFUSION or SO= JOURNAL OF CLINICAL ONCOLOGY or SO= AIDS or SO= AMERICAN JOURNAL OF PHYSICAL ANTHROPOLOGY)

#3

AD=INDIANA UNIV and (PY=1980 or PY=1982 or PY=1984 or PY=1986 or PY=1988 or PY=1990 or PY=1992 or PY=1994 or PY=1996 or PY=1998) not (SO=ANALYTICAL

BIOCHEMISTRY or SO= NEURON or SO= CLINICAL CHEMISTRY or SO= METHODS IN ENZYMOLOGY or SO=(AMERICAN JOURNAL OF OBSTETRICS AND GYNECOLOGY) or SO= (AIDS RESEARCH AND HUMAN RETROVIRUSES) or SO= JOURNAL OF MOLECULAR BIOLOGY or SO= (INFECTION AND IMMUNITY) or SO= JOURNAL OF THE AMERICAN COLLEGE OF CARDIOLOGY or SO= CHEST or SO=AMERICAN JOURNAL OF DISEASES OF CHILDREN or SO= ARCHIVES OF INTERNAL MEDICINE or SO= VIROLOGY or SO= (JOURNAL OF ACQUIRED IMMUNE DEFICIENCY SYNDROMES AND HUMAN RETROVIROLOGY) or SO= (CLINICAL ORTHOPAEDICS AND RELATED RESEARCH) or SO= (PHARMACOLOGY BIOCHEMISTRY AND BEHAVIOR) or SO= (PLASTIC AND RECONSTRUCTIVE SURGERY) or SO= CLINICAL PHARMACOLOGY & THERAPEUTICS or SO= DEVELOPMENTAL BIOLOGY or SO= (DRUG AND ALCOHOL DEPENDENCE) or SO= MEDICAL CARE or SO= AMERICAN JOURNAL OF PSYCHIATRY or SO= (FERTILITY AND STERILITY) or SO= TRANSPLANTATION PROCEEDINGS or SO= JOURNAL OF NEUROPHYSIOLOGY or SO= JOURNAL OF NUCLEAR MEDICINE or SO= TRANSPLANTATION or SO= AMERICAN JOURNAL OF PATHOLOGY or SO= CONTROLLED CLINICAL TRIALS or SO= CIRCULATION RESEARCH or SO= JOURNAL OF RHEUMATOLOGY or SO= SURGERY or SO= BULLETIN OF THE HISTORY OF MEDICINE or SO=CLINICAL INFECTIOUS DISEASES or SO=JOURNAL OF BACTERIOLOGY or SO= ARCHIVES OF NEUROLOGY or SO= CRITICAL CARE MEDICINE or SO= (CYTOGENETICS AND CELL GENETICS) or SO=INTERNATIONAL JOURNAL OF RADIATION ONCOLOGY BIOLOGY PHYSICS)

#4

AD=INDIANA UNIV and (PY=1980 or PY=1982 or PY=1984 or PY=1986 or PY=1988 or PY=1990 or PY=1992 or PY=1994 or PY=1996 or PY=1998) not (SO= JOURNAL OF EXPERIMENTAL MEDICINE or SO= PSYCHOPHARMACOLOGY or SO= ANNALS OF THORACIC SURGERY or SO=NEUROBIOLOGY OF AGING or SO= BIOLOGY OF REPRODUCTION or SO= HUMAN IMMUNOLOGY or SO= JOURNAL OF ORGANIC CHEMISTRY or SO= MAGNETIC RESONANCE IN MEDICINE or SO= (RETINA-THE JOURNAL OF RETINAL AND VITREOUS DISEASES) or SO= (JOURNAL OF TRAUMA-INJURY INFECTION AND CRITICAL CARE) or SO= PEDIATRIC INFECTIOUS DISEASE JOURNAL or SO= PROSTATE or SO= UROLOGY or SO= AMERICAN JOURNAL OF ROENTGENOLOGY or SO= ANNALS OF PLASTIC SURGERY or SO= (ARCHIVES OF BIOCHEMISTRY AND BIOPHYSICS) or SO= BIOCHIMICA ET BIOPHYSICA ACTA or SO= EPILEPSIA or SO= EUROPEAN JOURNAL OF PHARMACOLOGY or SO= GENE or SO= SEXUALLY TRANSMITTED DISEASES or SO= (AMERICAN JOURNAL OF RESPIRATORY AND CRITICAL CARE MEDICINE) or SO= (ARCHIVES OF PHYSICAL MEDICINE AND REHABILITATION) or SO= ARCHIVES OF GENERAL PSYCHIATRY or SO= ARCHIVES OF PEDIATRICS & ADOLESCENT MEDICINE or SO=ENVIRONMENTAL HEALTH PERSPECTIVES or SO= INTERNATIONAL JOURNAL OF EPIDEMIOLOGY or SO= (JOURNAL OF BONE AND JOINT SURGERY-AMERICAN VOLUME) or SO= JOURNAL OF COMPARATIVE NEUROLOGY or SO= MODERN PATHOLOGY or SO= HUMAN PATHOLOGY or SO= NEUROSCIENCE or SO= (OTOLARYNGOLOGY-HEAD AND NECK SURGERY) or SO= (ALCOHOLISM-CLINICAL AND EXPERIMENTAL RESEARCH) or SO=CARCINOGENESIS or SO= (CLINICAL IMMUNOLOGY AND IMMUNOPATHOLOGY) or SO= BIOLOGICAL PSYCHIATRY or SO=COLD SPRING HARBOR SYMPOSIA ON QUANTITATIVE BIOLOGY or SO= (JOURNAL OF OCCUPATIONAL AND ENVIRONMENTAL MEDICINE))

#5

AD=INDIANA UNIV and (PY=1980 or PY=1982 or PY=1984 or PY=1986 or PY=1988 or PY=1990 or PY=1992 or PY=1994 or PY=1996 or PY=1998) not (SO= GLYCOBIOLOGY or SO= SEMINARS IN ONCOLOGY or SO= SOUTHERN MEDICAL JOURNAL or SO= AMERICAN JOURNAL OF CLINICAL PATHOLOGY or SO= (ANESTHESIA AND ANALGESIA) or SO= ARCHIVES OF DERMATOLOGY or SO= EMBO JOURNAL or SO= LARYNGOSCOPE or SO= NEUROSCIENCE LETTERS or SO= AMERICAN JOURNAL OF SURGICAL PATHOLOGY or SO= ARCHIVES OF OTOLARYNGOLOGY-HEAD & NECK SURGERY or SO= CANCER EPIDEMIOLOGY BIOMARKERS & PREVENTION or SO= CELLULAR IMMUNOLOGY or SO= HUMAN GENETICS or SO= HUMAN MOLECULAR GENETICS or SO= JOURNAL OF PHYSIOLOGY-LONDON or SO= NATURE GENETICS or SO= PUBLIC HEALTH REPORTS or SO= TRENDS IN NEUROSCIENCES or SO= CLINICAL PEDIATRICS or SO= GYNECOLOGIC ONCOLOGY or SO= IEEE TRANSACTIONS ON BIOMEDICAL ENGINEERING or SO= JOURNAL OF CLINICAL PSYCHIATRY or SO= JOURNAL OF COMPUTER ASSISTED TOMOGRAPHY or SO= MEDICINE or SO= (PROTEINS-STRUCTURE FUNCTION AND GENETICS) or SO= AMERICAN HEART JOURNAL or SO= ANNALS OF SURGERY or SO= GENES & DEVELOPMENT or SO= JOURNAL OF GENERAL PHYSIOLOGY or SO= JOURNAL OF THE AMERICAN ACADEMY OF DERMATOLOGY or SO= MOLECULAR BRAIN RESEARCH or SO= MUSCLE & NERVE or SO= NATURE MEDICINE or SO= AMERICAN ZOOLOGIST or SO=ANNALS OF EMERGENCY MEDICINE or SO= (ANTIMICROBIAL AGENTS AND CHEMOTHERAPY) or SO= BIOMETRICS)

#6

AD=INDIANA UNIV and (PY=1980 or PY=1982 or PY=1984 or PY=1986 or PY=1988 or PY=1990 or PY=1992 or PY=1994 or PY=1996 or PY=1998) not (SO= INTERNATIONAL JOURNAL OF GYNECOLOGY & OBSTETRICS or SO= JOURNAL OF CLINICAL PHARMACOLOGY or SO= JOURNAL OF DENTAL RESEARCH or SO= JOURNAL OF REPRODUCTIVE MEDICINE or SO= (MOLECULAR AND BIOCHEMICAL PARASITOLOGY) or SO= AMERICAN JOURNAL OF NEURORADIOLOGY or SO= BRITISH JOURNAL OF OPHTHALMOLOGY or SO= EXPERIMENTAL BRAIN RESEARCH or SO= JOURNAL OF PEDIATRIC SURGERY or SO= JOURNAL OF PHYSICAL CHEMISTRY or SO= SCHIZOPHRENIA RESEARCH or SO= AMERICAN JOURNAL OF KIDNEY DISEASES or SO= ANATOMICAL RECORD or SO= EXPERIMENTAL NEUROLOGY or SO= FEBS LETTERS or SO= GERIATRICS or SO= INTERNATIONAL JOURNAL OF CANCER or SO= INVESTIGATIVE RADIOLOGY or SO= JOURNAL OF CLINICAL EPIDEMIOLOGY or SO= (JOURNAL OF DEVELOPMENTAL AND BEHAVIORAL PEDIATRICS) or SO=PSYCHOSOMATICS or SO= REVIEWS OF INFECTIOUS DISEASES or SO= (TOXICOLOGY AND APPLIED PHARMACOLOGY) or SO= UROLOGIC CLINICS OF NORTH AMERICA or SO=ACADEMIC MEDICINE or SO= ARTERIOSCLEROSIS or SO=(BEHAVIORAL AND BRAIN SCIENCES) or SO= (DEVELOPMENTAL MEDICINE AND CHILD NEUROLOGY) or SO= (ELECTROENCEPHALOGRAPHY AND CLINICAL NEUROPHYSIOLOGY) or SO=ENVIRONMENTAL MUTAGENESIS or SO= MOLECULAR IMMUNOLOGY or SO= NEUROSURGERY or SO= (PHOTOCHEMISTRY AND PHOTOBIOLOGY) or SO= STATISTICS IN MEDICINE or SO= SURVEY OF OPHTHALMOLOGY or SO= (AMERICAN JOURNAL OF RESPIRATORY CELL AND MOLECULAR BIOLOGY) or SO=

BIOCHEMICAL PHARMACOLOGY or SO= BULLETIN OF THE WORLD HEALTH ORGANIZATION or SO= EPIDEMIOLOGIC REVIEWS)

#7

AD=INDIANA UNIV and (PY=1980 or PY=1982 or PY=1984 or PY=1986 or PY=1988 or PY=1990 or PY=1992 or PY=1994 or PY=1996 or PY=1998) not (SO= EUROPEAN JOURNAL OF CELL BIOLOGY or SO= JOURNAL OF ADOLESCENT HEALTH or SO= JOURNAL OF ANDROLOGY or SO=LABORATORY ANIMAL SCIENCE or SO= PHYSIOLOGY & BEHAVIOR or SO= AMERICAN JOURNAL OF GASTROENTEROLOGY or SO= BEHAVIORAL NEUROSCIENCE or SO= (BRAIN AND LANGUAGE) or SO= DEVELOPMENTAL BRAIN RESEARCH or SO= GENETICS or SO= JOURNAL OF INVESTIGATIVE MEDICINE or SO= (JOURNAL OF MAGNETISM AND MAGNETIC MATERIALS) or SO= JOURNAL OF NEUROIMMUNOLOGY or SO= JOURNAL OF NUTRITION or SO= BRITISH JOURNAL OF PSYCHIATRY or SO= DIABETES CARE or SO= HOSPITAL PRACTICE or SO= (JOURNAL OF BONE AND MINERAL RESEARCH) or SO=(JOURNAL OF MOLECULAR AND CELLULAR CARDIOLOGY) or SO= (PACE-PACING AND CLINICAL ELECTROPHYSIOLOGY) or SO= AMERICAN JOURNAL OF INDUSTRIAL MEDICINE or SO= (ANNALS OF OTOLOGY RHINOLOGY AND LARYNGOLOGY) or SO= APPETITE or SO=BRAIN or SO= BRAIN RESEARCH BULLETIN or SO= BRITISH MEDICAL JOURNAL or SO= CIBA FOUNDATION SYMPOSIA or SO= (FREE RADICAL BIOLOGY AND MEDICINE) or SO= JOURNAL OF CELLULAR PHYSIOLOGY or SO= (JOURNAL OF NEUROPSYCHIATRY AND CLINICAL NEUROSCIENCES) or SO= LIFE SCIENCES or SO= MOLECULAR PHARMACOLOGY or SO= (OPHTHALMIC SURGERY AND LASERS) or SO= PSYCHOLOGICAL MEDICINE or SO= SURGICAL CLINICS OF NORTH AMERICA or SO=ACTA CYTOLOGICA or SO= (AMERICAN JOURNAL OF PHYSIOLOGY-HEART AND CIRCULATORY PHYSIOLOGY))

#8

AD=INDIANA UNIV and (PY=1980 or PY=1982 or PY=1984 or PY=1986 or PY=1988 or PY=1990 or PY=1992 or PY=1994 or PY=1996 or PY=1998) not (SO= AMERICAN JOURNAL OF SURGERY or SO= ARCHIVES OF PATHOLOGY & LABORATORY MEDICINE or SO= CANCER TREATMENT REPORTS or SO= EXPERIMENTAL EYE RESEARCH or SO= EXPERIMENTAL PARASITOLOGY or SO= JOURNAL OF GENERAL VIROLOGY or SO= NEUROTOXICOLOGY or SO= PAIN or SO= PERCEPTION & PSYCHOPHYSICS or SO=(APPLIED AND ENVIRONMENTAL MICROBIOLOGY) or SO= ARCHIVES OF SURGERY or SO= (CLINICAL AND DIAGNOSTIC LABORATORY IMMUNOLOGY) or SO= (CLINICAL AND EXPERIMENTAL ALLERGY) or SO= DEVELOPMENT or SO= EPILEPSY RESEARCH or SO= HEARING RESEARCH or SO= HEMATOLOGY-ONCOLOGY CLINICS OF NORTH AMERICA or SO= (HOSPITAL AND COMMUNITY PSYCHIATRY) or SO=JOURNAL OF HAND SURGERY-AMERICAN VOLUME or SO= (JOURNAL OF LABORATORY AND CLINICAL MEDICINE) or SO= (JOURNAL OF NEUROLOGY NEUROSURGERY AND PSYCHIATRY) or SO=(JOURNAL OF THE AMERICAN ACADEMY OF CHILD AND ADOLESCENT PSYCHIATRY) or SO=ONCOGENE or SO= PEDIATRIC CLINICS OF NORTH AMERICA or SO=PROTEIN SCIENCE or SO= TRENDS IN BIOCHEMICAL SCIENCES or SO= CARBOHYDRATE RESEARCH or SO= (CLINICAL AND EXPERIMENTAL IMMUNOLOGY) or SO= GENETIC EPIDEMIOLOGY or SO= INFECTIOUS DISEASES IN CLINICAL PRACTICE or SO= JOURNAL OF CELL SCIENCE or SO= JOURNAL OF LEUKOCYTE BIOLOGY or SO= JOURNAL OF PARASITOLOGY or SO= JOURNAL OF PEDIATRIC OPHTHALMOLOGY & STRABISMUS or SO= MACROMOLECULES or SO= SHOCK or SO= VACCINE or SO= VISION RESEARCH or SO=(ADVANCES IN EXPERIMENTAL MEDICINE AND BIOLOGY))

#9

AD=INDIANA UNIV and (PY=1980 or PY=1982 or PY=1984 or PY=1986 or PY=1988 or PY=1990 or PY=1992 or PY=1994 or PY=1996 or PY=1998) not (SO= AMERICAN INDUSTRIAL HYGIENE ASSOCIATION JOURNAL or SO= AMERICAN JOURNAL OF PHYSIOLOGY-CELL PHYSIOLOGY or SO= BRITISH JOURNAL OF HAEMATOLOGY or SO= GLYCOCONJUGATE JOURNAL or SO= JOURNAL OF MEDICINAL CHEMISTRY or SO= JOURNAL OF SURGICAL RESEARCH or SO= SURGERY GYNECOLOGY & OBSTETRICS or SO= (TRANSACTIONS OF THE ROYAL SOCIETY OF TROPICAL MEDICINE AND HYGIENE) or SO= TRENDS IN GENETICS or SO=AMERICAN FAMILY PHYSICIAN or SO= ANNALS OF BIOMEDICAL ENGINEERING or SO= CARDIOVASCULAR RESEARCH or SO= (DIGESTIVE DISEASES AND SCIENCES) or SO= GASTROINTESTINAL ENDOSCOPY or SO= HYPERTENSION or SO= (JOURNAL OF CLINICAL AND EXPERIMENTAL NEUROPSYCHOLOGY) or SO= JOURNAL OF NEUROSURGERY or SO= JOURNAL OF THE AMERICAN DIETETIC ASSOCIATION or SO= ORTHOPEDICS or SO= SPINE or SO= ANNUAL REVIEW OF NEUROSCIENCE or SO= BRITISH JOURNAL OF CANCER or SO= CLINICAL CANCER RESEARCH or SO= CURRENT BIOLOGY or SO= CURRENT EYE RESEARCH or SO= CURRENT OPINION IN OBSTETRICS & GYNECOLOGY or SO= FLUID PHASE EQUILIBRIA or SO= JOURNAL OF BIOMECHANICAL ENGINEERING-TRANSACTIONS OF THE ASME or SO= JOURNAL OF HUMAN EVOLUTION or SO=JOURNAL OF INHERITED METABOLIC DISEASE or SO= JOURNAL OF NEUROSCIENCE METHODS or SO= JOURNALS OF GERONTOLOGY or SO= MAMMALIAN GENOME or SO= NEUROCHEMICAL RESEARCH or SO= NURSING RESEARCH or SO= PHYSICAL THERAPY or SO= PREVENTIVE MEDICINE or SO= (AMERICAN JOURNAL OF PHYSIOLOGY-REGULATORY INTEGRATIVE AND COMPARATIVE PHYSIOLOGY))

#10

AD=INDIANA UNIV and (PY=1980 or PY=1982 or PY=1984 or PY=1986 or PY=1988 or PY=1990 or PY=1992 or PY=1994 or PY=1996 or PY=1998) not (SO= AMERICAN JOURNAL OF PREVENTIVE MEDICINE or SO= BEHAVIOURAL BRAIN RESEARCH or SO= BONE MARROW TRANSPLANTATION or SO= CLINICAL ELECTROENCEPHALOGRAPHY or SO= COMPREHENSIVE PSYCHIATRY or SO= EAR NOSE & THROAT JOURNAL or SO= HEART & LUNG or SO= IMMUNOLOGY or SO= JOURNAL OF THE EXPERIMENTAL ANALYSIS OF BEHAVIOR or SO= (MEDICAL AND PEDIATRIC ONCOLOGY) or SO= MEDICAL PHYSICS or SO= MUTATION RESEARCH or SO= SCHIZOPHRENIA BULLETIN or SO= ACTA PAEDIATRICA or SO= (AMERICAN JOURNAL OF PHYSIOLOGY-LUNG CELLULAR AND MOLECULAR PHYSIOLOGY) or SO=ANNALS OF ALLERGY or SO= BIOMETRIKA or SO= BRITISH JOURNAL OF SURGERY or SO= (CANCER CHEMOTHERAPY AND PHARMACOLOGY) or SO= CLINICAL GENETICS or SO= CURRENT OPINION IN CELL BIOLOGY or SO= CURRENT OPINION IN IMMUNOLOGY or SO= CURRENT OPINION IN INFECTIOUS DISEASES or SO= EXPERIMENTAL CELL RESEARCH or SO= (GRAEFES ARCHIVE FOR CLINICAL AND EXPERIMENTAL OPHTHALMOLOGY) or SO= INTERNATIONAL JOURNAL OF PSYCHIATRY IN MEDICINE or SO= JOURNAL OF MEDICAL GENETICS or SO= (JOURNAL OF TOXICOLOGY AND ENVIRONMENTAL HEALTH) or SO= JOURNAL OF WILDLIFE DISEASES or SO= LEUKEMIA or SO= MEDICAL CLINICS OF NORTH AMERICA or SO= NEUROREPORT or SO= PEDIATRIC ANNALS or SO= RADIATION RESEARCH or SO=ACTA OTO-LARYNGOLOGICA or SO= AMERICAN JOURNAL OF HOSPITAL PHARMACY or SO= AMERICAN JOURNAL OF INFECTION CONTROL)

#11

AD=INDIANA UNIV and (PY=1980 or PY=1982 or PY=1984 or PY=1986 or PY=1988 or PY=1990 or PY=1992 or PY=1994 or PY=1996 or PY=1998) not (SO= ANNUAL REVIEW OF BIOCHEMISTRY or SO= ANNUAL REVIEW OF PHYSIOLOGY or SO= ARCHIVES OF ENVIRONMENTAL HEALTH or SO= BIOPHYSICAL CHEMISTRY or SO= (BIOTECHNOLOGY AND BIOENGINEERING) or SO= (CANCER GENETICS AND CYTOGENETICS) or SO= (CATHETERIZATION AND CARDIOVASCULAR DIAGNOSIS) or SO= (CELL AND TISSUE KINETICS) or SO= CURRENT OPINION IN NEUROLOGY or SO= (DRUG METABOLISM AND DISPOSITION) or SO= EUROPEAN JOURNAL OF IMMUNOLOGY or SO= HORMONE RESEARCH or SO=(INFECTION CONTROL AND HOSPITAL EPIDEMIOLOGY) or SO= INTERNATIONAL ANESTHESIOLOGY CLINICS or SO= JOURNAL OF BIOMECHANICS or SO= JOURNAL OF CHILD NEUROLOGY or SO= JOURNAL OF CUTANEOUS PATHOLOGY or SO= JOURNAL OF EXPERIMENTAL ZOOLOGY or SO= JOURNAL OF GENERAL INTERNAL MEDICINE or SO= JOURNAL OF IMMUNOLOGICAL METHODS or SO= JOURNAL OF LIPID RESEARCH or SO= JOURNAL OF MEDICAL ENTOMOLOGY or SO= JOURNAL OF NEUROSCIENCE RESEARCH or SO= JOURNAL OF PROTOZOOLOGY or SO= JOURNAL OF THE AMERICAN VETERINARY MEDICAL ASSOCIATION or SO=JOURNAL OF THEORETICAL BIOLOGY or SO= M D COMPUTING or SO= MARYLAND STATE MEDICAL JOURNAL or SO= (METABOLISM-CLINICAL AND EXPERIMENTAL) or SO= MOLECULAR ENDOCRINOLOGY or SO= POSTGRADUATE MEDICINE or SO= PSYCHIATRY RESEARCH or SO= RADIOGRAPHICS or SO= (SOMATIC CELL AND MOLECULAR GENETICS) or SO= ACS SYMPOSIUM SERIES or SO=ACTA PSYCHIATRICA SCANDINAVICA or SO= AMERICAN JOURNAL OF ORTHOPSYCHIATRY)

#12

AD=INDIANA UNIV and (PY=1980 or PY=1982 or PY=1984 or PY=1986 or PY=1988 or PY=1990 or PY=1992 or PY=1994 PY=1996 or PY=1998) not (SO=AMERICAN SURGEON or SO= BEHAVIOURAL PHARMACOLOGY or SO= BIOPOLYMERS or SO= CLINICS IN PLASTIC SURGERY or SO=CONTRACEPTION or SO= DNA-A JOURNAL OF MOLECULAR & CELLULAR BIOLOGY or SO=INTERNATIONAL JOURNAL OF NEUROSCIENCE or SO= INTERNATIONAL JOURNAL OF THE ADDICTIONS or SO= JOURNAL OF ACQUIRED IMMUNE DEFICIENCY SYNDROMES or SO= JOURNAL OF CARDIOVASCULAR PHARMACOLOGY or SO= JOURNAL OF CHRONIC DISEASES or SO=JOURNAL OF MAMMALOGY or SO= JOURNAL OF MEMBRANE BIOLOGY or SO= JOURNAL OF NEUROCYTOLOGY or SO= (JOURNAL OF PAIN AND SYMPTOM MANAGEMENT) or SO=(JOURNAL OF PEDIATRIC GASTROENTEROLOGY AND NUTRITION) or SO= JOURNAL OF STRUCTURAL BIOLOGY or SO= (JOURNAL OF THE HISTORY OF MEDICINE AND ALLIED SCIENCES) or SO= JOURNAL OF VASCULAR SURGERY or SO= MEDICAL DECISION MAKING or SO= MEDICAL JOURNAL OF AUSTRALIA or SO= PFLUGERS ARCHIV-EUROPEAN JOURNAL OF PHYSIOLOGY or SO=PHARMACOTHERAPY or SO= PHILOSOPHICAL TRANSACTIONS OF THE ROYAL SOCIETY OF LONDON SERIES B-BIOLOGICAL SCIENCES or SO= PIGMENT CELL RESEARCH or SO= PROGRESS IN BRAIN RESEARCH or SO= SLEEP or SO= SURGICAL NEUROLOGY or SO= SYNAPSE or SO=TISSUE ANTIGENS or SO= ULTRASONICS ADDICTION or SO=AICHE JOURNAL or SO=AMERICAN JOURNAL OF HEMATOLOGY or SO= AMERICAN JOURNAL OF NURSING or SO=AMERICAN JOURNAL OF THE MEDICAL SCIENCES or SO= ANNUAL REVIEW OF PUBLIC HEALTH)

#13

AD=INDIANA UNIV and (PY=1980 or PY=1982 or PY=1984 or PY=1986 or PY=1988 or PY=1990 or PY=1992 or PY=1994 or PY=1996 or PY=1998) not (SO=ANTICANCER RESEARCH or SO=ANTIVIRAL RESEARCH or SO=BIOCHEMICAL SOCIETY TRANSACTIONS or SO=BIOLOGICAL BULLETIN or SO=BRITISH JOURNAL OF PHARMACOLOGY or SO=CYTOMETRY or SO=DEVELOPMENTAL PSYCHOBIOLOGY or SO=DRUGS or SO=(ENVIRONMENTAL TOXICOLOGY AND CHEMISTRY) or SO=EUROPEAN JOURNAL OF BIOCHEMISTRY or SO=EUROPEAN JOURNAL OF CLINICAL NUTRITION or SO=FOOT & ANKLE INTERNATIONAL or SO=JOURNAL OF BIOLOGICAL CHEMISTRY or SO=BIOCHEMICAL JOURNAL or SO=BIO* or SO=MOLECULAR* or SO=CELLULAR* or SO=CLINICAL* or SO=CELL* or SO=MEDICINE* or SO=NEURO* or SO=BRAIN* or SO=TRENDS IN PHARMA* or SO=PSYCH* or SO=NATURE or SO=SCIENCE) not (TS=BIOCHEMISTRY or TS=NEUROSCIENCES)

#14

AD=INDIANA UNIV and (PY=1980 or PY=1982 or PY=1984 or PY=1986 or PY=1988 or PY=1990 or PY=1992 or PY=1994 or PY=1996 or PY=1998) not (TS=PHARMACOLOGY or TS= PHARMACY or TS= NEUROLOGY or TS= IMMUNOLOGY or  TS= BIOLOGY or  TS= MEDICINE or  TS= ONCOLOGY or  TS= SURGERY or  TS= GENETICS or TS= HEREDITY or  TS= ENDOCRINOLOGY or TS= METABOLISM or  TS= TOXICOLOGY or  TS= RADIOLOGY or TS= MEDICAL IMAGING or  TS= PEDIATRICS or  TS= OPHTHALMOLOGY or  TS= PSYCHIATRY or  TS= PATHOLOGY or  TS= OCCUPATIONAL HEALTH or  TS= CARDIAC or TS= CARDIOVASCULAR or  TS= PHYSIOLOGY or  TS= BIOTECHNOLOGY or  TS= MICROBIOLOGY or  TS= ZOOLOGY or TS= OBSTETRICS or TS= GYNECOLOGY or  TS= HEMATOLOGY or  TS= BIOPHYSICS or  TS= PERIPHERAL VASCULAR DISEASE or  TS= BEHAVIORAL SCIENCES or  TS= RESPIRATORY SYSTEM or  TS= ALLERGY or TS=DIETETICS or  TS= UROLOGY)

#15

AD=INDIANA UNIV and (PY=1980 or PY=1982 or PY=1984 or PY=1986 or PY=1988 or PY=1990 or PY=1992 or PY=1994 or PY=1996 or PY=1998) not (TS= GASTROENTEROLOGY or  TS= HEPATOLOGY or  TS= ORTHOPEDICS or  TS= PSYCHOLOGY or  TS= DERMATOLOGY or  TS= PARASITOLOGY or  TS= OTORHINOLARYNGOLOGY or  TS= GERIATRICS or  TS= GERONTOLOGY or  TS=

BIOLOGICAL or TS= RHEUMATOLOGY or TS= ANESTHESIOLOGY or TS= MEDICAL LABORATORY or TS= VIROLOGY or TS= DENTISTRY or TS= ORAL SURGERY or TS= HEALTH or TS= PLANT SCIENCES or TS= SOCIAL SCIENCES or TS= SPORT SCIENCES or TS= MICROSCOPY or TS= NEPHROLOGY or TS= VETERINARY SCIENCES)

#16

AD=INDIANA UNIV and (PY=1980 or PY=1982 or PY=1984 or PY=1986 or PY=1988 or PY=1990 or PY=1992 or PY=1994 or PY=1996 or PY=1998) not (TS= PALEONTOLOGY or TS= CLINICAL or TS= SOCIAL ISSUES or TS= ENTOMOLOGY or TS= ECONOMICS or TS= INFECTIOUS DISEASES or TS= MEDICAL INFORMATICS or TS= ORNITHOLOGY or TS= AGRICULTURE or TS= SOIL SCIENCE or TS= NURSING or TS= AGRICULTURE or TS= MULTIDISCIPLINARY or TS= FOOD SCIENCE or TS= FORESTRY or TS= LEGAL or TS= MYCOLOGY or TS= ANDROLOGY or TS=LINGUISTICS or TS= SUBSTANCE ABUSE or TS= AGRICULTURE or TS= DAIRY or TS=ANIMAL SCIENCE or TS= ANTHROPOLOGY or TS= AGRONOMY or TS= BIODIVERSITY CONSERVATION or TS= BUSINESS or TS= ERGONOMICS or TS= FISHERIES or TS= REHABILITATION)

Followed by the command:

#1 and #2 and #3 and #4 and #5 and #6 and #7 and #8 and #9 and #10 and #11 and #12 and #13 and #14 and #15 and #16

After the selection of the set of publications formed by combining the 16 separate searches, the set was weeded by hand. A typical example of excluded Subject Categories is:

PLANT SCIENCES, BIOCHEMISTRY & MOLECULAR BIOLOGY, ZOOLOGY, ENTOMOLOGY, BIOLOGY, GENETICS & HEREDITY, ECONOMICS, AGRICULTURAL ECONOMICS & POLICY, CELL BIOLOGY, ENDOCRINOLOGY & METABOLISM, BEHAVIORAL SCIENCES, PHYSIOLOGY, FOOD SCIENCE & TECHNOLOGY, MARINE & FRESHWATER BIOLOGY, OPHTHALMOLOGY, AGRICULTURE, MULTIDISCIPLINARY, FORESTRY, RADIOLOGY, NUCLEAR MEDICINE & MEDICAL IMAGING, NEUROSCIENCES, TOXICOLOGY, NUTRITION & DIETETICS, AGRICULTURE, SOIL SCIENCE, PUBLIC, ENVIRONMENTAL & OCCUPATIONAL HEALTH, MICROBIOLOGY, PHARMACOLOGY & PHARMACY, BIOTECHNOLOGY & APPLIED MICROBIOLOGY, BIOCHEMICAL RESEARCH METHODS, AGRONOMY, DEVELOPMENTAL BIOLOGY, BIODIVERSITY CONSERVATION, MEDICINE, LEGAL, ORNITHOLOGY, TRANSPORTATION, PALEONTOLOGY, MEDICINE, RESEARCH & EXPERIMENTAL, SOCIAL SCIENCES, INTERDISCIPLINARY, PSYCHOLOGY, MICROSCOPY, IMMUNOLOGY, MYCOLOGY, ONCOLOGY, HEMATOLOGY, MEDICINE, GENERAL & INTERNAL, ANATOMY & MORPHOLOGY, PERIPHERAL VASCULAR DISEASE, PATHOLOGY, SOCIAL ISSUES, SUBSTANCE ABUSE, VETERINARY SCIENCES, ENGINEERING, BIOMEDICAL, PSYCHOLOGY, AGRICULTURE, DAIRY & ANIMAL SCIENCE, EVOLUTIONARY BIOLOGY, REPRODUCTIVE BIOLOGY, GERIATRICS & GERONTOLOGY, OBSTETRICS & GYNECOLOGY, PARASITOLOGY, VIROLOGY, BIOPHYSICS, LAW, PEDIATRICS, FISHERIES, HORTICULTURE, MUSIC, NURSING, HEALTH CARE SCIENCES & SERVICES

Reference Table; These impact indices were produced using publications weeded for bio-medical and other unrelated topics, and including even years from 1980 to 1990. Note that the impact indices for NRAO, STSCI, and UC Berkeley are higher than when calculated for the 9 years 1990-1998 as shown in Table 5 in the main text, while NOAO remains approximately the same.

| Institute | N | h-index | h(m) |
|---|---|---|---|
| Astrophysics Institutes | | | |
| UC Berkeley, Dept Astr | 1076 | 111 | 6.80 |
| CALTECH, Dept Astr | 657 | 68 | 6.57 |
| Institute of Astronomy, Cambridge UK | 1923 | 125 | 6.07 |
| STScI* | 1410 | 105 | 5.77 |
| Harvard Smithsonian Center for Astrophy | 2703 | 130 | 5.51 |
| NOAO (all sites)** | 917 | 82 | 5.37 |
| NRAO (all sites) | 1106 | 86 | 5.21 |
| | | | |
| Space Physics, Geophysics, Planetary Institutes | | | |
| NCAR, Boulder CO | 2675 | 135 | 5.75 |
| Scripps Inst | 2597 | 123 | 5.30 |
| Space Science Lab, UC Berkeley | 802 | 76 | 5.24 |
| IGPP, all sites | 1744 | 103 | 5.20 |
| Lunar & Planetary Lab, Tucson, AS | 755 | 62 | 4.38 |
| | | | |

- * No publications for 1980.
- ** No publications for 1980 or 1982.